\documentclass[prd,twocolumn,preprintnumbers]{revtex4}
\usepackage{eucal} 
\usepackage{bm}
\usepackage{graphicx}
\newcommand{\be}{\begin{eqnarray}}
\newcommand{\ee}{\end{eqnarray}}
\newcommand{\beq}{\begin{equation}}
\newcommand{\eeq}{\end{equation}}
\begin{document}
\preprint{JLAB-THY-08-810}
\title{Exclusive electroproduction of strange mesons with JLab 
12 GeV\footnote{Prepared for the CLAS 12 RICH Detector Workshop, 
January 28--29, 2008, Jefferson Lab, Newport News, VA, USA; \\
\url{http://conferences.jlab.org/clas12/}.}
}
\author{M.~Strikman}
\affiliation{Department of Physics, Pennsylvania State University,
University Park, PA 16802, USA}
\author{C.~Weiss}
\affiliation{Theory Center, Jefferson Lab, Newport News, VA 23606, USA}
\begin{abstract}
We summarize the physics topics which can be addressed by measurements 
of high--$Q^2$ exclusive electroproduction of strange mesons, 
$\gamma^\ast N \rightarrow \phi N, \, K^\ast \Lambda, \, K\Lambda, \,
K\Sigma$, at Jefferson Lab with 11 GeV beam energy. The proposed
investigations are aimed both at exploring the reaction mechanism
(dominance of point--like configurations) and extracting information
about baryon structure from the data (generalized parton distributions, 
or GPDs). They include 
(a) probing the $t$--dependence of the nucleon's gluon GPD (transverse 
spatial distribution of gluons) in $\phi$ meson production; 
(b) separating the nucleon helicity--flip and nonflip quark GPDs 
in $K^\ast \Lambda$ production with measurement of the $\Lambda$
recoil polarization; (c) probing strangeness
polarization in the nucleon in $K\Lambda$ and $K\Sigma$ production.
These studies rely only on the analysis of cross section ratios, 
which are less affected by the theoretical uncertainties of 
present GPD--based calculations than absolute cross sections.
\end{abstract}
\maketitle
\section{Introduction}
Exclusive electroproduction of mesons, $\gamma^\ast N \rightarrow 
M N'$, at energies above the resonance region, 
$W > 2 \, \text{GeV}$, and low momentum transfer to the 
nucleon, $|t| \lesssim 1 \, \text{GeV}^2$, offers many interesting 
opportunities for studying strong interaction dynamics 
as well as baryon and meson structure at variable resolution scale, 
defined by the photon virtuality, $Q^2$. At low virtualities,
$Q^2 \lesssim 1\, \text{GeV}^2$, such reactions are conventionally 
described in terms of hadronic
fluctuations of the virtual photon and their interaction with the
target. At high virtualities, $Q^2 \gg 1\, \text{GeV}^2$, the
production process becomes effectively point--like and can be 
described as the interaction of the virtual photon with partonic
degrees of freedom in the target. In the asymptotic regime, 
a QCD factorization theorem \cite{Collins:1996fb} allows one 
to calculate the electroproduction
amplitudes in terms of the generalized parton distributions (GPDs) 
of the target (more precisely, of the $N \rightarrow N'$ transition) 
and the distribution amplitudes (DAs) of the produced meson $M$ --- 
universal characteristics of the quark and gluon structure of the hadrons, 
which are probed also in other hard scattering processes such as 
inclusive DIS, elastic scattering, and $e^+e^-$ annihilation
(see Refs.~\cite{Goeke:2001tz,Diehl:2003ny,Belitsky:2005qn} for reviews). 

The analysis of such processes at high $Q^2$ typically proceeds
in two stages. In the first stage, one tries to ascertain the approach
to the point--like regime by testing certain qualitative features
which do not depend on the specific form of the GPDs/DAs and the details 
of the hard scattering process (examples of such tests will be given below). 
In the second stage, one may then try to extract quantitative information 
about the GPDs and/or DAs by comparing suitable process observables 
with corresponding theoretical predictions calculated assuming 
factorization. Of particular interest is 
the fact that a given meson channel probes specific quantum numbers 
of the $N \rightarrow N'$ transition GPD, making it possible to separate 
different spin/flavor components and perform comparative studies 
of related channels. While the practical applicability of the factorized
approximation to meson production at $Q^2 \sim \text{few GeV}^2$ is far 
from established, and most likely requires substantial sub-asymptotic 
corrections (``higher twist''), the study of such processes can in 
principle contribute much valuable information to the GPD analysis.

In this article we attempt to summarize which questions could be 
addressed by measurements of high--$Q^2$ exclusive electroproduction 
of strange mesons, in the kinematic region accessible with the 
12 GeV Upgrade of Jefferson Lab. Our survey covers three main channels, 
each of which has its own distinctive physical interest and potential 
use in the GPD analysis: 
\be
\gamma^\ast N &\rightarrow& \phi N , 
\label{phi} \\
\gamma^\ast p &\rightarrow& K^{\ast +} \Lambda, 
\label{kstar_lambda} \\
\gamma^\ast p &\rightarrow& K^+ \Lambda, \, K^+ \Sigma^0, \, K^0 \Sigma^+ .
\label{k_lambda} 
\ee
$\phi$ meson production, Eq.~(\ref{phi}), allows one to investigate
the approach to the point--like regime through the transverse momentum
transfer dependence of the cross section, and to study the $t$--dependence 
of the nucleon's gluon GPD and obtain information about the transverse 
spatial distribution of gluons (Sec.~\ref{sec:phi}). 
Production of $K^\ast \Lambda$ with measurement of the transverse
recoil polarization, Eq.~(\ref{kstar_lambda}), 
can be used to separate the helicity--flip and nonflip 
components of the $N \rightarrow \Lambda$ transition GPDs, complementing
measurements with a transversely polarized target. Even more information
on the helicity structure could be obtained by combining target and 
recoil polarization. In these experiments
$\sigma_L$ and $\sigma_T$ can effectively be separated by analyzing the
angular distribution of the vector meson decay and relying on
$s$--channel helicity conservation; this method avoids comparing data 
at different beam energies, which is costly and limits the kinematic 
reach of the experiments. Further information 
about the flavor structure of quark GPDs can be gained from measurements
of the ratio of $K^{\ast +}\Lambda$ and $\rho^+ n$ production 
cross sections (Sec.~\ref{sec:kstar}). 
Finally, $K\Lambda$ and $K\Sigma$ production 
Eq.~(\ref{k_lambda}), are sensitive 
to the polarized $N \rightarrow N'$ transition 
GPDs $\tilde H$ and $\tilde E$, 
which by $SU(3)$ flavor symmetry can be related to the strangeness 
polarization in the nucleon (Sec.~\ref{sec:kaon}). 

An essential point is that the investigations described here rely on the 
analysis of cross section ratios rather than absolute cross sections.
As explained in Sec.~\ref{sec:status}, present GPD--based calculations 
of absolute meson production cross sections at JLab energies 
suffer from considerable
theoretical uncertainties. Measurements of cross section ratios,
in which these uncertainties cancel at least partly, presently
seem to be the best option for utilizing meson production for 
the GPD analysis at JLab 12 GeV. One may hope, of course, that
progress in the theory and phenomenology of exclusive meson 
production would eventually allow us to analyze also absolute cross 
section observables in these processes.

In strange meson electroproduction at high $Q^2$ the cross sections for 
processes with transition to decuplet baryon transitions are generally 
of the same order as those for octet baryons. This makes it possible to 
use hard exclusive processes of the type discussed here as a new tool in 
resonance spectroscopy. While not the main focus of this summary, 
we comment on this interesting option in Secs.~\ref{sec:status}
and \ref{sec:summary}.
\section{Theoretical status of high--$Q^2$ meson production}
\label{sec:status}
\begin{figure}[t]
\includegraphics[width=.30\textwidth]{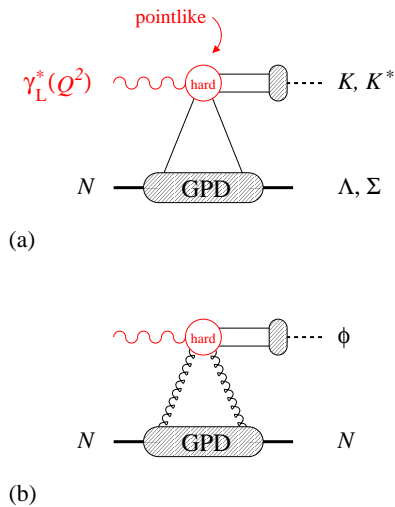}
\caption[]{QCD factorization in exclusive electroproduction
of strange mesons. $K$ and $K^\ast$ production (a) probe 
nonsinglet quark GPDs, $\phi$ production (b) mostly the 
gluon GPD.}
\label{fig:fact}
\end{figure}

The theoretical basis for the analysis of processes such as 
Eqs.~(\ref{phi})--(\ref{k_lambda}) at high $Q^2$ is the QCD factorization 
theorem for hard exclusive meson electroproduction \cite{Collins:1996fb}.
At $Q^2 \gg R_{\text{hadron}}^{-2}$, the meson is produced in a 
configuration much smaller than its typical hadronic size.
The amplitude for longitudinal ($L$) photon polarization can be
factorized into a hard scattering process, calculable in perturbative QCD,
the $N \rightarrow N'$ transition GPD, describing the emission/absorption 
of the active parton (quark or gluon) by the nucleon, and the DA of the 
produced meson, describing the hadronization of the outgoing 
$q\bar q$ pair (see Fig.~\ref{fig:fact}). Factorization in these
processes is closely related to the phenomenon of color transparency ---
the fact that color--singlet configurations of size 
$r \ll R_{\text{hadron}}$ interact weakly with hadronic matter; 
this connection is seen most clearly when following the space--time 
evolution of the reaction in the target rest frame \cite{Frankfurt:2005mc}.
%
%
%

%
%
\begin{figure}[b]
\includegraphics[width=.28\textwidth]{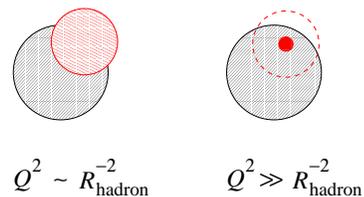}
\caption[]{The transverse spatial structure of the interaction region
in meson production at $Q^2 \sim R_{\text{hadron}}^{-2}$ 
(left) and $Q^2 \gg R_{\text{hadron}}^{-2}$ (right). 
The photon virtuality determines the effective transverse size 
of the configurations in which the meson is produced. 
Experimentally, the transverse size of the interaction
region is reflected in the $t$--slope (more precisely, the 
$\bm{\Delta}_\perp^2$--slope) of the differential cross section.}
\label{fig:discs}
\end{figure}
There are many indications for substantial corrections to the asymptotic
reaction mechanism at momentum transfers $Q^2 \sim \text{few GeV}^2$, 
originating from production of the meson in configurations of 
finite size $1/Q \ll r \lesssim R_{\text{hadron}}$. In the kinematic
region of $x_B \ll 10^{-1}$, clear evidence for such corrections comes 
from the $Q^2$--dependence of the 
$t$--slopes of several vector meson production channels 
($\rho^0, \phi, J/\psi$) measured at HERA 
(see Ref.~\cite{Frankfurt:2005mc} for details).  The $t$--slope,
which at small $x_B$ coincides with the slope in the
transverse momentum transfer to the target, $\bm{\Delta}_\perp^2$
(see Sec.~\ref{sec:phi}), measures the total transverse size of the 
interaction region, as determined by the size of the relevant 
configurations of the target and the produced meson, 
and becomes $Q^2$--independent 
in the asymptotic regime where the meson is produced in a point--like 
configuration (see Fig.~\ref{fig:discs}). 
The data show that $\rho^0$ and $\phi$ slope
still decrease up to $Q^2 \approx 10 \, \text{GeV}^2$, indicating 
that there is still a significant contribution from finite--size 
configurations \cite{Frankfurt:2005mc}. It is worth emphasizing that
this observation could not possibly be explained by higher--order
QCD corrections and must be attributed to finite--size effects. 
(Next--to--leading order QCD corrections to meson production 
amplitudes were studied in Refs.~\cite{Diehl:2007hd}.)
Further evidence for finite--size effects,
more pertinent to the kinematic region of $x_B \gtrsim 10^{-1}$ 
relevant to JLab, comes from the extensive studies of the 
pion form factor at high $Q^2$, whose asymptotic behavior is governed by 
a hard scattering mechanism closely related to high--$Q^2$ exclusive 
meson production (see Refs.~\cite{Belitsky:2005qn,Radyushkin:1998vb}
for reviews). In fact, the charged pion/kaon production amplitude 
contains a ``pole term'' governed by the pion/kaon form factor
(see Sec.~\ref{sec:kaon}), and the findings about finite--size
effects in the pion form factor directly impact on the GPD--based 
description of meson production.

In theoretical calculations of meson production cross sections 
based on QCD factorization one faces several questions:
(a) how to model the GPDs; (b) how to treat the hard scattering process
(choice of scale in $\alpha_s$, higher--order corrections); 
(c) how to consistently combine contributions from meson 
production in small--size and large--size configurations.
While in theory these are distinct questions which can be 
discussed separately, in practice the issues are closely related, 
implying that the approximations made in the treatment of one will 
generally influence the conclusions one draws about the others.
The situation is reasonably well under control in vector meson production
at collider energies (HERA, EIC), where the dominant gluon GPD can be 
reconstructed at $t = 0$ from the usual gluon density in a 
well--controlled approximation, and the finite size of the produced
meson (intrinsic transverse momentum) can be incorporated 
phenomenologically in the dipole picture in space--time, justified by 
the large coherence length of the process, 
$l_{\text{coh}} \sim 1/(2 M_N x_B) \gg 1 \, \text{fm}$ 
\cite{Frankfurt:1995jw}. (Another approach to finite--size effects 
at small $x_B$, based on intrinsic transverse momentum, 
was pursued in Ref.~\cite{Goloskokov:2005sd}.)
In the kinematics of fixed--target experiments 
(HERMES, JLab) the situation is generally more complicated.
Present model calculations of absolute meson production cross sections 
in this region show considerable uncertainty (see \textit{e.g.}\ 
Refs.~\cite{Vanderhaeghen:1999xj,Diehl:2005gn}). 
Progress can be expected from further 
theoretical studies of the reaction mechanism (space--time picture, 
real vs.\ imaginary part of the amplitude), 
as well as from fully differential cross section measurements 
($t$-- and $W$--dependences for given $Q^2$; $L/T$ separation
and other response functions, polarization observables) with JLab 12 GeV. 

Given the present uncertainties in GPD--based calculations of
absolute meson production cross sections, a reasonable approach is to 
concentrate on the analysis of cross section ratios in which these 
uncertainties cancel at least to some extent. Such ``ratio observables'' 
can be used either to test certain qualitative predictions of the approach 
to the point--like regime, or to extract specific information 
about the GPDs. Examples are:
\begin{itemize}
\item The $t$--dependence of the cross section and its change 
with $Q^2$, which measures the transverse size of the interaction
region and the $t$--dependence of the GPDs;
\item Target and recoil polarization asymmetries, which can be
used to separate the nucleon helicity components of the GPDs;
\item Beam single--spin and beam + target double--spin 
asymmetries, which probe asymptotically subleading amplitudes 
with transverse ($T$) virtual photon polarization and offer clues
about the reaction mechanism in this sector;
\item Ratios of cross sections of similar channels, \textit{e.g.}\
$K^{\ast +}/\rho^+, K^0/\pi^0$, etc., which test the flavor structure
of the nucleon GPDs.
\end{itemize}
Specific examples of the use of such observables in strange meson
production will be described below.

%
%
\begin{table}
\[
\renewcommand{\arraystretch}{1.7}
\begin{array}{ll} 
\hline
\rho^0 p & \frac{1}{\sqrt{2}} [2u + d] 
         + \frac{1}{\sqrt{2}} [2 \bar{u} + \bar{d}] + \frac{9}{4}\, g \\
\omega p & \frac{1}{\sqrt{2}} [2u - d] 
         + \frac{1}{\sqrt{2}} [2\bar{u} - \bar{d}] 
         + \frac{3}{4}\, g \\
\phi p & - [s + \bar s] + \frac{3}{4}\, g \\
\rho^+ n & 2[u - d] - [\bar{u} - \bar{d}] \\
K^{*+} \Lambda & 
          -\frac{2}{\sqrt{6}} [2u - d - s] \\
       &  + \frac{1}{\sqrt{6}} [2 \bar{u} - \bar{d} - \bar{s}] \\
K^{*+}\Sigma^0 & -\frac{2}{\sqrt{2}} [d - s] 
              + \frac{1}{\sqrt{2}} [\bar{d} - \bar{s}] \\
K^{*0}\Sigma^+ & [d - s] + [\bar{d} - \bar{s}]
\rule[-0.8em]{0pt}{1em} \\
\hline
\pi^+ n & 2 [\Delta u - \Delta d] 
          + [\Delta \bar{u} - \Delta \bar{d}] \\
\pi^0 p & \frac{1}{\sqrt{2}} [2\Delta u + \Delta d] 
        - \frac{1}{\sqrt{2}} [2\Delta \bar{u} + \Delta \bar{d} ] \\
K^+ \Lambda & - \frac{2}{\sqrt{6}} [2\Delta u - \Delta d - \Delta s] \\
            &
               - \frac{1}{\sqrt{6}} [2\Delta \bar{u} - \Delta \bar{d} 
                  - \Delta \bar{s}] \\
K^+ \Sigma^0 & - \frac{2}{\sqrt{2}} [\Delta d - \Delta s]
  - \frac{1}{\sqrt{2}} [\Delta \bar{d} - \Delta \bar{s}] \\
K^0\Sigma^+ & [\Delta d - \Delta s] 
                - [\Delta \bar{d} - \Delta \bar{s} ] \\
\hline
\end{array}
\renewcommand{\arraystretch}{1}
\]
\caption{Spin/flavor combination of GPDs entering in the 
amplitudes of hard exclusive vector (top) and pseudoscalar (bottom) 
meson production with proton target, $\gamma^\ast p \rightarrow M N'$,
assuming $SU(3)$ flavor 
symmetry \cite{Goeke:2001tz,Diehl:2003ny,Diehl:2005gn}. 
The $p \rightarrow N'$ transition GPDs have been converted to
flavor--diagonal GPDs in the proton using relations such as
Eq.~(\ref{su3}). Vector meson production probes the ``unpolarized''
GPDs $H$ and $E$ (denoted symbolically by $u, d, s, g$), 
pseudoscalar meson production the ``polarized'' GPDs $\tilde H$ and 
$\tilde E$ (denoted by $\Delta u, \Delta d, \Delta s$).}
\label{table:su3}
\end{table}
An important theoretical tool in describing the electroproduction of
strange mesons is $SU(3)$ flavor symmetry, which has been extensively
tested and used in the analysis of strong reactions and weak decays 
of strange particles (see Ref.~\cite{Guzey:2005vz} for a recent review).
$SU(3)$ symmetry allows one to relate the nucleon to octet 
hyperon transition matrix element of a quark bilinear operator to a linear
combination of diagonal matrix elements in the proton, \textit{e.g.} 
\beq
\langle \Lambda | \bar s u | p \rangle 
\; = \; - \frac{1}{\sqrt{6}} \left[ 
2 \langle p | \bar u u | p \rangle - \langle p | \bar d d | p \rangle - 
\langle p | \bar s s | p \rangle \right] ;
\label{su3}
\eeq
in this way one can relate the $N\rightarrow \Lambda$ transition GPDs
to the usual flavor--diagonal GPDs in the proton. Table~\ref{table:su3}
lists the spin/flavor components of the proton GPD accessible in 
meson electroproduction with octet baryon final states.
Note that $SU(3)$ symmetry breaking effects might be quite different in 
the valence ($q - \bar q$) and sea quark ($\bar q$) distributions 
(see \textit{e.g.}\ Ref.~\cite{Lichtenstadt:1995xk}); the two
are probed in different combinations in the various exclusive channels.

Exclusive meson production can also be studied in processes where the 
nucleon undergoes a transition to a decuplet baryon, for example
the $\Delta$. In such processes the hard scattering process can be
thought of as an operator inducing the $N \rightarrow 
\text{resonance}$ transition. When combined with strangeness production,
such processes offer interesting possibilities to investigate
resonance structure with quantum numbers not accessible in usual
photo/electroexcitation. An important theoretical tool in the analysis
of such processes is the large--$N_c$ limit of QCD, in which baryons
can be described as classical field configurations (solitons), 
and octet and decuplet baryons appear as different rotational states 
of the same underlying object \cite{Frankfurt:1999xe}.
\section{Transverse gluon imaging in $\phi$ meson production}
\label{sec:phi}
Electroproduction of $\phi$ mesons is in many ways the simplest 
hard exclusive process accessible at JLab with 6 and 11 GeV beam energy.
Like $\rho^0$ and $\omega$ production, $\phi$ production proceeds by 
exchange of vacuum quantum numbers between the target and the projectile,
and has a reasonable cross section 
($\sim 10^1 \, \text{nb}$ for $W = 2-3 \, \text{GeV}$
and $Q^2 = 1-2 \, \text{GeV}^2$ 
\cite{Cassel:1981sx,Lukashin:2001sh,Santoro:2008ai}). 
However, contrary to the light vector mesons, in $\phi$ production 
quark exchange is suppressed, making it a clean probe of the gluon 
field in the nucleon. Calculations of the $\phi$ production
cross section in the leading--twist approximation, using standard
parametrizations of the strange quark and gluon GPDs at a low scale, 
suggest that $\phi$ production is dominated by the gluon GPD 
even at JLab energies (however, the leading--twist approximation
cannot reliably predict the absolute cross section) \cite{Diehl:2005gn};
\textit{cf.}\ also the analysis of Ref.~\cite{Goloskokov:2007nt} at 
smaller values of $x_B$. Experimentally, the analysis of $\phi$ meson 
production is aided by the fact that one can infer the polarization of the 
produced vector meson from the measurement of its decay angular 
distribution and use $s$--channel helicity conservation to 
extract $\sigma_L$, eliminating the need for explicit $L/T$ separation 
by way of measurements at different beam energies (Rosenbluth separation).

The most interesting ratio observable in $\phi$ production
is the dependence of the differential cross section on the transverse 
momentum transfer to the target, $\bm{\Delta}_\perp$,
\beq
\frac{d\sigma_L /dt \, (\bm{\Delta}_\perp^2 ) \phantom{= 0} \,}
{d\sigma_L /dt \, (\bm{\Delta}_\perp^2 = 0)} 
\;\; = \;\; \text{function}(\bm{\Delta}_\perp^2, x_B, Q^2) ,
\label{t_dep}
\eeq
and its change with $x_B$ and $Q^2$. Here we have in mind 
values of the order $|\bm{\Delta}_\perp| \lesssim 1 \, \text{GeV}$, which 
carry the information about the transverse structure of the interaction
region. The transverse momentum transfer
is related to the invariant momentum transfer $t$ by
\beq
t - t_{\text{min}} \;\; = \;\; \frac{-\bm{\Delta}_\perp^2}{1 - \xi^2} , 
\label{t_from_Delta_T}
\eeq
where $\xi$ is $1/2$ times the relative longitudinal momentum loss of 
the target, which in the case of $Q^2 \gg \text{(masses)}^2$ 
is given by
\beq
\xi \;\; = \;\; \frac{x_B}{2 - x_B} .
\eeq
QCD factorization predicts that the ratio (\ref{t_dep}) become 
independent of $Q^2$ at fixed $x_B$. In the subasymptotic 
regime, the change of the $\bm{\Delta}_\perp^2$--dependence with $Q^2$ 
reveals the decrease of the transverse size of the interaction region 
with $Q^2$, and thus provides information about the effective transverse 
size of the $s\bar s$ pair forming the $\phi$ meson 
(see Fig.~\ref{fig:discs}). At $x_B \ll 10^{-1}$ (HERA, EIC) one 
has $\xi \ll 1$; moreover, one can neglect $t_{\text{min}}$ in 
Eq.~(\ref{t_from_Delta_T}), so that $t \approx -\bm{\Delta}_\perp^2$,
and the $\bm{\Delta}_\perp^2$ dependence can be identified with the
$t$--dependence of the cross section. At $x_B \gtrsim 10^{-1}$ (JLab)
one must distinguish between the $\bm{\Delta}_\perp^2$ and $t$ dependences 
when studying the decrease of the transverse size of the interaction
region with increasing $Q^2$. Also, at $Q^2 \sim \text{few} \,\text{GeV}^2$
it might be necessary to use the exact expression for the longitudinal
momentum transfer $\xi$, including power--suppressed terms of the
form $\text{(mass)}^2/Q^2$. Observing the expected broadening of
the the $\bm{\Delta}_\perp^2$ distribution Eq.~(\ref{t_dep}) 
with increasing $Q^2$ and its eventual 
stabilization (corresponding to shrinkage of the transverse size of the 
interaction region, \textit{cf.}\ Fig.~\ref{fig:discs}) 
affords a simple quantitative test of the approach 
to the pointlike regime, independent of specific GPD models.
We emphasize again that it is the $\bm{\Delta}_\perp^2$--dependence at
small values, corresponding to $|t - t_{\text{min}}| \lesssim 1 \, 
\text{GeV}^2$, which is of interest here, not at values of the 
order of $|t - t_{\text{min}}| \sim Q^2$.

For sufficiently large $Q^2$, when the ratio Eq.~(\ref{t_dep})
has become approximately $Q^2$--independent, it can be compared
with the ratio of cross sections calculated in terms of GPDs.
In the leading--twist approximation,
\beq
\frac{d\sigma_L /dt \, (\bm{\Delta}_\perp^2) \phantom{= 0} \,}
{d\sigma_L /dt \, (\bm{\Delta}_\perp^2 = 0)} 
\;\; = \;\; \frac{|\mathcal{F}_g|^2 (\xi, t)
\phantom{= 0} \,}{|\mathcal{F}_g|^2 
(\xi, t_{\text{min}})} ,
\label{rat_lt}
\eeq
where $\bm{\Delta}_\perp^2$ and $t$ are related by 
Eq.~(\ref{t_from_Delta_T}), and \cite{Diehl:2003ny}
\be 
|\mathcal{F}_g|^2 (\xi, t) &\equiv& 
(1 - \xi^2) |\mathcal{H}_g|^2 
\nonumber \\
&-& (\xi^2 + t/4M^2) |\mathcal{E}_g|^2 
\nonumber \\
&-& 2\xi^2 \text{Re}(\mathcal{E}_g^\ast \mathcal{H}_g) .
\label{amp2}
\ee
Here $\mathcal{H}_g$ and $\mathcal{E}_g$ are the (complex) 
leading--twist amplitudes associated with the gluon GPDs,
\be
\mathcal{H}_g &=& e_s^2 \;
\int_{-1}^1 dx \; H_g (x, \xi, t) 
\nonumber \\
&\times& \left( \frac{1}{\xi - x + i0} 
- \frac{1}{\xi + x + i0} \right) \;\;\; \textit{etc}.
\label{H_g}
\ee
Note that the theoretical prediction for the ratio Eq.~(\ref{rat_lt})
is independent of the form of the DA of the produced $\phi$ meson 
and the details of the treatment of the hard scattering process.
The GPDs here are taken at an effective scale $Q^2_{\text{eff}} < Q^2$,
determined by the effective transverse size of the $s\bar s$ pair
in the production process. 
The ratio Eq.~(\ref{rat_lt}) can be used to constrain the 
$t$--dependence of the gluon GPDs, and thus the transverse
spatial distribution of gluons in the nucleon, in a largely 
model--independent manner, free of the uncertainties of 
present absolute cross section calculations in the GPD approach.

At $x_B \ll 0.1$ the $\phi$ production cross section is mostly due to 
the proton helicity--conserving amplitude $\mathcal{H}_g$ [the contribution 
of the proton helicity--flip amplitude $\mathcal{E}_g$ in Eq.~(\ref{amp2}) 
is suppressed], and the amplitude $\mathcal{H}_g$ is dominated by
its imaginary part, in which the gluon GPD is sampled at
$x = \xi \approx x_B/2$. 
In this kinematic region Eq.~(\ref{rat_lt}) takes the simple form
\beq
\frac{d\sigma_L /dt \, (t) \phantom{= 0} \, }
{d\sigma_L /dt \, (t = 0)} 
\;\; = \;\; 
\frac{H_g^2 (x = \xi, \xi, t) \phantom{= 0} \;\; }
{H_g^2 (x = \xi, \xi, t = 0)} ,
\label{t_dep_GPD}
\eeq
and the $t$--dependence of the cross section can directly be 
interpreted in terms of the $t$--dependence of the gluon GPD.
In this approximation the data for $J/\psi, \phi$ and $\rho^0$
production were analyzed in Refs.~\cite{Frankfurt:2005mc,Frankfurt:2002ka}.
The phenomenological interpretation of the data in this region and our 
theoretical understanding of the $t$--dependence of the gluon GPD
and its change with $x$ are summarized in Ref.~\cite{Frankfurt:2005mc}. 

At $x_B \gtrsim 10^{-1}$, the analysis of the ratio Eq.~(\ref{rat_lt})
should include the helicity--flip gluon GPD $E_g$ and the presence
of a possibly sizable real part of the leading--twist amplitudes.
In particular, a real part of the amplitude can arise from the D--term 
in the gluon GPDs, which is not constrained by the forward limit,
and whose magnitude is largely unknown \cite{Polyakov:1999gs}. 
We note that a sizable gluonic D--term would influence also the 
leading--twist predictions for $\rho^0$ production (see 
Ref.~\cite{Guidal:2007cw} for a discussion of the preliminary CLAS data).

%
%
\begin{figure}
\includegraphics[width=.45\textwidth]{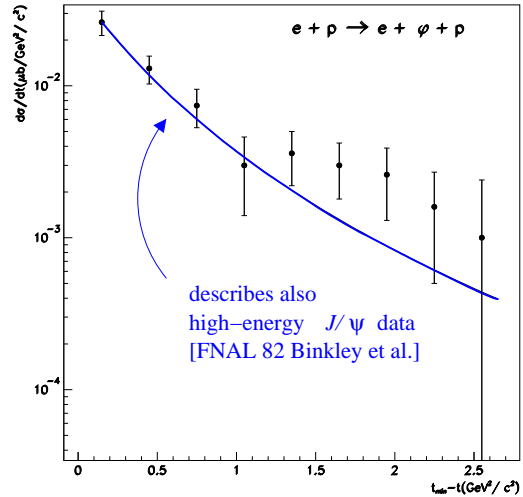}
\caption[]{The differential cross section of exclusive $\phi$
meson production, $\gamma^\ast p \rightarrow \phi p$, as a function
of $t$, as measured by JLab CLAS \cite{Lukashin:2001sh}. 
The curve shows a fit by a $t$--dependence of the form 
$d\sigma/dt \propto (1 - t/1.0 \, {\rm GeV}^2)^{-4}$,
corresponding to a dipole form of the $t$--dependence of 
the gluon GPD, which describes well the $J/\psi$ photoproduction
data from the FNAL E401 / E458 experiments \cite{Binkley:1981kv}.}
\label{fig:phi}
\end{figure}
Figure~\ref{fig:phi} shows the $t$--dependence of the 
$\phi$ production cross section measured in the JLab CLAS 
experiment \cite{Lukashin:2001sh} (new data were presented
recently in Ref.~\cite{Santoro:2008ai}). The curve shows a fit by a 
$t$--dependence of the form $d\sigma/dt \propto 
(1 - t/1.0 \, {\rm GeV}^2)^{-4}$, corresponding to a dipole form 
of the $t$--dependence of $H_g$ in the simplified expression 
Eq.~(\ref{t_dep_GPD}). This form is theoretically motivated
by the analogy of the large--$x$ two--gluon form factor with the 
nucleon axial form factor \cite{Frankfurt:2002ka} and describes well
the $J/\psi$ photoproduction data from the FNAL E401 / E458 
experiments at higher energies and a larger scale 
$Q^2_{\text{eff}}$ \cite{Binkley:1981kv}. One sees
that it fits well the $t$--dependence of the CLAS data for 
$|t| \lesssim 1 \, \text{GeV}^2$. This is very encouraging,
and supports the universal $\bm{\Delta}_\perp^2$ dependence of these
processes implied by the approach to the pointlike regime
(assuming the two probe comparable $x$--values in the gluon GPD).

In the leading--twist approximation, the Fourier transform of the 
$\bm{\Delta}_\perp$--dependence of the cross section ratio
Eq.~(\ref{t_dep}) is related to the impact parameter dependence 
of the GPDs \cite{Burkardt:2002hr,Diehl:2002he}. Provided that
$Q^2$ is sufficiently large to ensure dominance of point--like 
meson configurations, this can be used to take transverse images of 
the relevant partonic configurations in the target in these processes.
The interpretation of these images is particularly simple at 
$x_B \ll 0.1$, where one can approximate the gluon GPDs by the 
``diagonal'' one ($\xi = 0$), and the coordinate variable $\bm{b}$
measures the distance of the active quark from a fixed ($x$--independent)
transverse center of the nucleon. A more general interpretation, 
valid in the case $\xi \neq 0$ relevant to JLab experiments, 
has been described in Ref.~\cite{Diehl:2002he}. 
Generally, one expects the transverse
size of the relevant partonic configurations in meson production 
to decrease with increasing $x_B$ (see also the discussion in
Refs.~\cite{Frankfurt:2002ka,Frankfurt:2005mc}). Preliminary results
on the $x_B$--dependence of the $t$--slope in $\rho^0$ production
have recently been reported by the CLAS Collaboration 
at JLab \cite{Guidal:2007cw}.

Another interesting observable is the ratio of the $\phi$ to the
$\rho^0$ and $\omega$ production cross sections. If all vector 
meson production amplitudes were dominated by gluon exchange,
the cross section ratio would be independent of $x_B$, 
\beq
\frac{\sigma_L (\gamma^\ast p \rightarrow \phi p)}
{\sigma_L (\gamma^\ast p \rightarrow \rho^0 p)}
\;\; = \;\; \text{const.}
\hspace{2em} (\text{gluon exch.})
\eeq
Any $x_B$ dependence of the ratio therefore indicates the presence of 
quark exchange. This test would be particularly instructive
at moderately small $x_B$ ($10^{-2} \lesssim x_B \lesssim 10^{-1}$), 
where the amplitudes are expected to be largely gluon--dominated 
but singlet quark exchange still makes a noticeable contribution
\cite{Goloskokov:2007nt}. In the valence quark region 
($x_B \sim 0.2 - 0.5$), where quark
exchange is expected to be large, the interpretation of a 
nontrivial $x_B$ dependence of the ratio is more model--dependent, 
as the different quark distributions could in principle ``conspire'' 
to produce a similar $x_B$--dependence as gluon exchange over a 
limited range. Still, the ratio is a good observable for testing
the relative importance of singlet quark and gluon exchange 
at JLab energies. An interesting cross--check would be to 
analyze also the ratio of $\rho^+ n$ to $\rho^0 p$ production
cross sections; the former is due to nonsinglet quark exchange only, 
the latter involves both quark and gluon exchange.

Electroproduction of $\phi$ mesons could also be used to probe
the light--quark component ($\bar u u, \bar d d$) of the $\phi$ meson,
which arises due to $\omega\phi$ and $\rho\phi$ mixing. Specifically, 
$\phi$ meson production with $p \rightarrow \Delta^0$ transition,
\beq
\gamma^\ast p \;\; \rightarrow \phi \Delta^0 ,
\eeq
requires $I = 1$ and thus light--quark exchange in the $t$--channel,
and couples primarily to the light--quark component of the $\phi$.
In this example the $\Delta^0$ acts as a ``filter'' for the quantum
numbers transferred to the produced meson. The analysis of this 
process could also incorporate information on isospin violation
in the $\phi$ meson from the $\phi \rightarrow \omega \pi^0$
decay measured recently in $e^+e^-$ 
annihilation \cite{Ambrosino:2007wf,Li:2008xm}.
\section{Helicity--flip GPD $E$ from recoil polarization
in $K^\ast \Lambda$ production}
\label{sec:kstar}
The ``unpolarized'' quark GPDs in the nucleon come in the form
of two functions, $H$ and $E$, corresponding to the Dirac and Pauli
form factors in the matrix element of the vector current, and related 
to the possibility of helicity--conserving and helicity--flipping 
transitions between the nucleon states. The Pauli form factor--type 
GPD $E$ is of special significance for nucleon structure. 
Contrary to the Dirac form factor--type GPD $H$, which at zero momentum 
transfer coincides with the usual unpolarized quark density, 
the $x$--dependence of $E$ is essentially unknown. Only its 
first moment can be inferred from the nucleon's anomalous magnetic moment,
and it shows that the distribution is sizable and likely to play a 
significant role in hard exclusive amplitudes. In the impact parameter
representation, the function $E$ describes the distortion of the 
quarks' longitudinal motion by transverse polarization of the 
nucleon state \cite{Burkardt:2002hr}. It is also needed as an 
ingredient to the angular momentum sum rule \cite{Ji:1996nm}. 

Extracting information about the Pauli form--factor type GPD $E$ from 
experimental data presents a major challenge. In the $N(e, e'\gamma)N$
cross section due to DVCS and Bethe--Heitler interference the 
contribution of $E$ is suppressed for a proton target, and only
experiments with neutron (\textit{i.e.}, nuclear) targets offer reasonable
sensitivity. More direct access to $E$ is possible through leading--twist
polarization observables in vector meson production. The transverse target 
spin dependence of the $\gamma^\ast_L p \rightarrow \rho^0_L p$ cross 
section is caused by the interference of nucleon helicity--flip and
nonflip amplitudes and given by a term linear in the GPD $E$ 
\cite{Goeke:2001tz}. The same term can be accessed by measuring
the polarization of the recoiling baryon in scattering from an unpolarized
target. Such measurements are possible in $\gamma^\ast_L p \rightarrow 
K^{\ast +} \Lambda$, taking advantage of the ``self--analyzing'' nature 
of the $\Lambda$ --- the fact that the orientation of the 
$\Lambda \rightarrow p \pi^-$ decay plane is almost perfectly correlated 
with the $\Lambda$ spin. Such measurements have been widely discussed
in connection with semi--inclusive particle production.

More precisely, the differential cross sections for the production of 
longitudinally polarized vector mesons, $\gamma^\ast_L(q) + N(p) 
\rightarrow V_L (q') + N'(p')$, are of the form
\be
\sigma_L (\text{target pol.}) &=& \sigma_0 + (\bm{n} \bm{S}) 
\, \sigma_1, \\
\sigma_L (\text{recoil pol.}) &=& \sigma_0 + (\bm{n} \bm{S}') 
\, \sigma'_1,
\ee
where $\bm{S}$ and $\bm{S}'$ are the target and recoil spin
vectors, and $\bm{n} = \bm{q}' \times \bm{q} / |\bm{q}' \times \bm{q}|$
is a transverse vector normal to the scattering plane. The general 
amplitude structure of the process implies that $\sigma'_1 = \sigma_1$ 
for production of natural parity ($1^-$) vector mesons ($\rho, K^\ast$). 
In the leading--twist approximation, the relative polarization asymmetries
are given by (\textit{cf.}\ Refs.~\cite{Goeke:2001tz,Diehl:2003ny})
\be
\frac{\sigma_1}{\sigma_0} &=& \frac{\sigma'_1}{\sigma_0} 
\label{asym_gpd}
\\
&=& \frac{|\bm{\Delta}_\perp | \, 
\text{Im}(\mathcal{E}^\ast \mathcal{H})}
{(1 - \xi^2) |\mathcal{H}|^2 - (\xi^2 + t/4M^2) |\mathcal{E}|^2 
- 2\xi^2 \text{Re}(\mathcal{E}^\ast \mathcal{H})} ,
\nonumber
\ee
where $\cal{H}$ and $\cal{E}$ are the complex leading--twist amplitudes 
proportional to the GPDs, as defined in Ref.~\cite{Diehl:2003ny}
[see also Eq.~(\ref{H_g})]. In order to extract this cross section ratio 
from recoil polarization
data one defines the angle, $\beta$, of the recoil spin relative 
to the scattering plane, $(\bm{n}\bm{S}') = |\bm{S}'| \sin\beta$
(with sign as specified by the above definition of $\bm{n}$),
and calculates the angular asymmetry of the cross section
\be
A({\text{recoil pol.}}) 
&\equiv& \frac{\displaystyle \int_0^{\pi/2} \!\! d\beta \; 
\sigma_L(\beta) \; - \; \int_{-\pi/2}^0 \!\! d\beta \; \sigma_L (\beta)}
{\displaystyle \int_0^{\pi/2} \!\! d\beta \; \sigma_L(\beta) 
\; + \; \int_{-\pi/2}^0 \!\! d\beta \; \sigma_L (\beta)}
\nonumber \\
&=& \frac{2 |\bm{S}'| \sigma_1}{\pi \sigma_0} .
\label{A_R}
\ee

When measuring the asymmetry Eq.~(\ref{A_R}) in 
$\gamma^\ast_L p \rightarrow K^{\ast +} \Lambda$, 
the GPDs probed [\textit{cf.}\ Eq.~(\ref{asym_gpd})] are the 
$p \rightarrow \Lambda$ transition GPDs. They are of interest
in themselves, containing useful information about hyperon 
structure. Alternatively, one can use $SU(3)$ symmetry to relate them 
to the flavor--diagonal GPDs in the proton, Eq.~(\ref{su3}), and in this
way extract information about the elusive proton GPD $E$. Such 
analysis should eventually include $SU(3)$ breaking both in the 
meson distribution amplitude (where it induces an asymmetry of
the meson distribution amplitude) as well as in the GPDs.
We note that, as in $\phi$ and $\rho^0$ production, the cross section for 
longitudinal photon polarization can be isolated by measuring the
$K^{\ast} \rightarrow K\pi$ decay and relying on $s$--channel helicity
conservation.

Measurements of recoil polarization in 
$\gamma^\ast_L p \rightarrow K^{\ast +} \Lambda$ could 
also be done with a polarized target. This combination in principle
would allow one to measure the cross sections corresponding to 
individual helicity amplitudes, making it possible to separate 
$H$ and $E$ without relying on interference effects.

Another interesting observable is the ratio of $\rho^+ n$ 
and $K^{\ast +} \Lambda$ production cross sections. The $\rho^+ n$ 
channel involves nonsinglet quark exchange only, and GPD--based 
calculations of the cross section are free from the uncertainties 
in the relative strength of gluon and singlet quark exchange 
affecting the $\rho^0 p$ channel \cite{Vanderhaeghen:1999xj,Diehl:2005gn}. 
While not fully model--independent, one can expect the ratio of 
$\rho^+ n$ and $K^{\ast +} \Lambda$ production cross sections to be 
more reliably described by GPD--based calculations in the leading--twist 
approximation than the absolute cross sections. 

A rough estimate of the expected $K^{\ast +} \Lambda / \rho^+ n$
cross section ratio can be obtained if we neglect the differences in the 
final--state masses, use the flavor structure of the production amplitudes 
as given by $SU(3)$ symmetry (see Table~\ref{table:su3}), assume 
that $H_u = 2 H_d$ and $E_u = 2 E_d$, and neglect the contributions 
from $H_s$ and $E_s$. With these approximations we obtain
\beq
\frac{\sigma_L (\gamma^\ast p \rightarrow K^{\ast +} \Lambda)}
{\sigma_L (\gamma^\ast p \rightarrow \rho^+ n)}
\;\; \approx \;\; \frac{3}{2} ,
\eeq
showing that both are of the same order. More detailed calculations
have been reported in Ref.~\cite{Diehl:2005gn}. 

Experiments aiming to study $K^\ast\Lambda$ production must take
into account that the cross section for the $K^\ast\Sigma^0$ channel
is likely to be of comparable magnitude. It will be necessary to 
separate the two channels, as the $\Sigma^0$ decays to $\Lambda$ via emission
of a low--energy photon. This caveat applies also to $K\Lambda$ and
$K^\ast\Sigma^0$ production as discussed in Sec.~\ref{sec:kaon}.
\section{Strangeness polarization in $K\Lambda$ and $K\Sigma$ production}
\label{sec:kaon}
Pseudoscalar meson production at high $Q^2$ probes the ``polarized''
GPDs $\tilde H$ and $\tilde E$, whose first moments are given by  
the axial and pseudoscalar form factors of the axial vector current
operator. At zero momentum transfer the GPD $\tilde H$ coincides with
the usual polarized quark densities in the nucleon. In this sense,
pseudoscalar meson production experiments can probe the spin structure 
of the nucleon without using target 
polarization \cite{Mankiewicz:1998kg,Frankfurt:1999fp}.

\begin{figure}
\includegraphics[width=.33\textwidth]{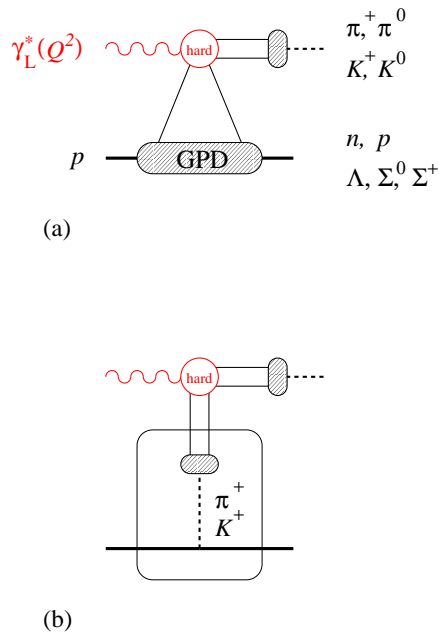}
\caption[]{(a) QCD factorization in pseudoscalar meson production. (b)
``Pole'' contribution to the GPD $\tilde E$ in $\pi^+$ and $K^+$
production. This contribution to the amplitude is equivalent to 
the virtual photon scattering from a $\pi^+ / K^+$ emitted by the proton.}
\label{fig:kaon}
\end{figure}

A special feature of the pseudoscalar GPD $\tilde E$ is that it contains
a term corresponding to $t$--channel exchange of pseudoscalar mesons 
(see Fig.~\ref{fig:kaon}), analogous to the ``pole term'' in the 
pseudoscalar form factor \cite{Mankiewicz:1998kg,Frankfurt:1999fp}.
In the context of meson production this term corresponds to the process
in which a $\pi^+$ or $K^+$ is emitted by the nucleon (with $|t| \sim 
M_{\pi}^2, M_{K}^2$) and interacts with the electromagnetic probe as 
a whole, via its EM form factor; in fact, this process is the basis of 
measurements of the $\pi^+$ and $K^+$ form factor in electroproduction
from the nucleon. Calculations based on the chiral quark--soliton model
of the nucleon \cite{Penttinen:1999th} show that the $\pi^+$ pole 
term dominates the isovector GPD $\tilde E$ at small $t$
and is largely responsible for the $\pi^+$ electroproduction cross section 
at $|t| \sim M_\pi^2$. The $K^+$ pole term in the $p \rightarrow \Lambda$
GPD is less prominent (because the pole at $t = M_K^2$ is further removed 
from the physical region $t < t_{\text{min}} < 0$), but it still
contributes significantly to the $K^+$ production cross section
\cite{Diehl:2005gn}. From the point of view of $SU(3)$ symmetry the
pole term represents a strong symmetry--breaking effect, as its 
strength depends on the pole position determined by the $\pi/K$ mass,
and unbroken symmetry would imply $M_\pi = M_K$.

%
%
In order to access the spin structure of the nucleon as encoded
in the GPD $\tilde H$, and to be able to apply $SU(3)$ symmetry,
one needs to separate the ``pole'' and ``non--pole'' contributions
in pseudoscalar meson production. In the case of pion production
this is in principle possible by comparing the $\pi^+ n$ channel
with $\pi^0 p$, where the pole contribution is exactly zero because 
the $\pi^0$ is a $C$--parity eigenstate and does not have a single--photon 
coupling. However, the two channels involve the isoscalar 
($\Delta u + \Delta d$) and isovector ($\Delta u - \Delta d$) 
GPDs with different coefficients, so that no fully
model--independent analysis is possible. In the case of kaon
production, the ``pole'' and ``non--pole'' contributions could
be separated by comparing the $K^+\Sigma^0$ channel, which has
a pole contribution, with the $K^0\Sigma^+$ channel, where the
pole contribution is not exactly zero but very small, at the level
of $SU(3)$ breaking mass effects in the $K^0$ wave function.
An interesting aspect of this comparison is that, when assuming
$SU(3)$ symmetry, the $K^+\Sigma^0$ and $K^0\Sigma^+$ GPDs 
involve the same nonsinglet quark flavor combination 
($\Delta d - \Delta s$) of the GPDs, so that the remaining
``non--pole'' terms in the GPDs can be compared in a 
relatively model--independent way.

Also of interest would be the comparison of $K$ production with
proton and neutron (\textit{i.e.}, nuclear) targets. If the $K^+$
production amplitude were dominated by the pole term, one would
expect that the ratio
\beq
\frac{\sigma_L (\gamma^\ast n \rightarrow K^+ \Sigma^-)}
{\sigma_L (\gamma^\ast p \rightarrow K^+ \Sigma^0)}
\eeq
be independent of $x_B$. If, however, the amplitude were mainly
due to the non--pole term, one would expect significant $x_B$--dependence,
as the $x$--dependence of the non--pole part of the GPD is governed by 
the polarized $u$--quark distribution, which is very different for proton 
and neutron targets (the $u$--quark distribution in the neutron is equal 
to the $d$--quark distribution in the proton). This study could be
complemented by measurement of the corresponding ratio for
$K^0$ production,
\beq
\frac{\sigma_L (\gamma^\ast n \rightarrow K^0 \Sigma^0)}
{\sigma_L (\gamma^\ast p \rightarrow K^0 \Sigma^=)} .
\eeq
Because the pole term in $K^0$ production is suppressed,
this ratio is expected to show much stronger $x_B$ dependence
than the one for $K^+$.
\section{Summary and Outlook}
\label{sec:summary}
Ratio observables play a crucial role in the analysis of exclusive 
meson production experiments at JLab with 6 and 
11 GeV beam energy. By studying suitable ratios of cross sections 
of similar channels, or of cross sections in the same channel at 
different kinematic points, one can perform numerous model--independent 
tests of the reaction mechanism, investigate the approach to the
point--like regime, and extract limited but specific information about 
the nucleon GPDs.

The extension to strangeness significantly enhances the reach of the 
exclusive meson production program at JLab with 11 GeV beam energy. 
$\phi$ meson production is an exceptionally clean channel; it probes 
the gluonic structure of the nucleon already at JLab energies, 
and the physical interpretation is closely related that of diffractive 
vector meson production at higher energies. $K^\ast \Lambda$ production
offers a unique way of studying the helicity structure
of the nucleon GPDs through recoil polarization measurements.
Experimentally, the requirements of particle identification and
energy/momentum resolution in the processes described here vary greatly 
between the channels and need to be discussed case--by--case.

As already noted, in high--$Q^2$ meson electroproduction processes
of the kind discussed here the cross sections for transition to decuplet 
baryons are generally of the same order as those for octet baryons. 
At 11 GeV beam energy, the produced baryon resonance and its decay 
products will be well--separated from the forward--going meson,
so that hadronic final--state interactions are small and both
can be studied as independent systems. Many interesting possibilities 
are connected with the use of hard exclusive processes for the purposes 
of resonance spectroscopy. Here the meson production process can be
thought of as an operator inducing the $N \rightarrow$ resonance 
transition. An interesting aspect is that these operators can probe 
flavor/parity quantum numbers not accessible in usual 
photo/electroexcitation experiments. Also, because the transition 
operators are nonlocal in the longitudinal direction (they remove 
a quark at one point and re-insert it at a different longitudinal 
position), they probe the orbital structure of the resonance wave 
function in a way different from local currents. The application 
of high--$Q^2$ exclusive processes to baryon resonance studies 
certainly merits further study.

Exclusive meson production could also be studied at higher CM energies
($W \gtrsim 10 \, \text{GeV}$) with a future electron--ion collider (EIC).
``Diffractive'' channels ($J/\psi, \phi, \rho^0, \text{DVCS}$), which at
high $Q^2$ probe the gluon and singlet quark GPDs, have cross sections 
which increase with energy and are relatively straightforward 
to measure. ``Nondiffractive'' channels ($\pi, K, \rho^+, K^\ast, \ldots$), 
which probe the spin/flavor/charge nonsinglet quark GPDs, have cross 
sections which decrease with energy and therefore require much higher
luminosity. The experimental feasibility of nondiffractive meson 
production with an EIC is presently being studied; first results
have been reported in Ref.~\cite{EIC}.
\begin{acknowledgments}
Notice: Authored by Jefferson Science Associates, LLC under U.S.\ DOE
Contract No.~DE-AC05-06OR23177. The U.S.\ Government retains a
non--exclusive, paid--up, irrevocable, world--wide license to publish or
reproduce this manuscript for U.S.\ Government purposes.
\end{acknowledgments}
\end{document}